\def\year{2020}\relax
\documentclass[letterpaper]{article} 
\usepackage{aaai20}  
\usepackage{times}  
\usepackage{helvet} 
\usepackage{courier}  
\usepackage[hyphens]{url}  
\usepackage{graphicx} 
\usepackage{graphicx}  
\usepackage{xcolor}
\usepackage{amssymb}
\usepackage{booktabs} 
\usepackage{amsmath}

\title{Towards an Integrative Educational Recommender \\ for Lifelong Learners 
}
\author{
\Large \textbf{Sahan Bulathwela, Mar\'ia P\'erez-Ortiz, Emine Yilmaz and John Shawe-Taylor}\\ 
Department of Computer Science, University College London \\
Gower Street, London WC1E 6BT, UK\\ 
\{m.bulathwela, maria.perez, emine.yilmaz, j.shawe-taylor\}@ucl.ac.uk  
}
 
\begin{document}

 \maketitle

\begin{abstract}

One of the most ambitious use cases of computer-assisted learning is to build a recommendation system for lifelong learning. Most recommender algorithms exploit similarities between content and users, overseeing the necessity to leverage sensible learning trajectories for the learner. Lifelong learning thus presents unique challenges, requiring scalable and transparent models that can account for learner knowledge and content novelty simultaneously, while also retaining accurate learners representations for long periods of time. We attempt to build a novel educational recommender, that relies on an integrative approach combining multiple drivers of learners engagement. Our first step towards this goal is TrueLearn, which models content novelty and background knowledge of learners and achieves promising performance while retaining a human interpretable learner model.


\end{abstract}

\section{Introduction}
As the world population grows, more innovative approaches should be sought to provide  high quality lifelong learning education opportunities to people of diverse cultures, languages, age groups and backgrounds.
Machine learning now promises to provide such benefits of personalised teaching to anyone in the world cost effectively.
 Since learner engagement is a prerequisite for achieving impactful learning outcomes~\cite{lan2017behavior}, we attempt to build a recommender system that models different drivers of engagement, 
assisting learners on their \emph{personal learning trajectory} to achieve their learning goals. Our approach differs from previous work in that it (i) incorporates different drivers of engagement such as resource quality, novelty, learner knowledge and interests; (ii) matches learners to useful and \emph{engaging fragments} of knowledge, as opposed to lengthy full resources; and (iii) supports a multi-lingual and multi-modal collection of learning resources.


\section{Related Work} \label{topic:related_work}

Conventional recommendation systems that exist today mainly focus on exploiting user interests. On the contrary, educational recommenders face different challenges as a successful educational recommender ought to satisfy additional functionalities, that stem from attempting to bring learners closer to their goals effectively. Some 
additional features worth mentioning are accounting for the novelty of materials~\cite{Drachsler:edurec} and identifying sensible learning trajectories. 
Although handcrafting learning trajectories~\cite{bauman2018recommending} is an option, such an approach is highly domain specific and lacks scalability. Similarly, handcrafting the \emph{Knowledge Components (KCs)} (or topics/concepts) present in a resource also poses similar drawbacks, which motivate the need for an automatic, domain-agnostic entity linking algorithm.
Incorporating these additional features to the system envisages i) detecting learners interests and goals, as these can significantly affect their motivation \cite{Salehi2014}; ii) detecting the current knowledge state of learners, the topics covered in a resource and the prerequisites necessary for benefiting from a learning material \cite{bauman2018recommending}; iii) recommending novel and relevant educational resources; and iv) accounting for how different content features of a resource impact how engaging a resource is
\cite{quality_features,Guo_vid_prod}.


The majority of work in adaptive educational systems builds on Item Response Theory (IRT)~\cite{Rasch1960,Pelanek2017} and Knowledge Tracing (KT)~\cite{yudelson2013individualized} that focus on estimating learner's knowledge for a narrow set of skills based on test answers. The work focusing on modelling a wide spectrum of skills over longer periods of time, which is our main focus, is surprisingly scarse.  

While excelling on the personalisation front, there are other features that are often overlooked when designing educational recommendation systems. We design our system with these features in mind: 
(i) {Cross-modality} (e.g. video, text, audio etc.) and (ii) {cross-linguality} are vital to identifying and recommending educational resources across different modalities and languages. In a lifelong learning setting, these two features will allow matching learning resources to the most suitable learners that come from various backgrounds. (iii) {Transparency} empowers learners by building trust while supporting the learner's metacognition processes, such as planning, monitoring and reflection~\cite{Bull2016}. (iv) {Scalability} and (v) {data efficiency} allows maintaining the states of large masses of learners over longer periods of time while making the best use of available user signals, such as implicit engagement~\cite{Salehi2014}.

\begin{figure}[!tbp]
 \centering
 \includegraphics[width=\columnwidth]{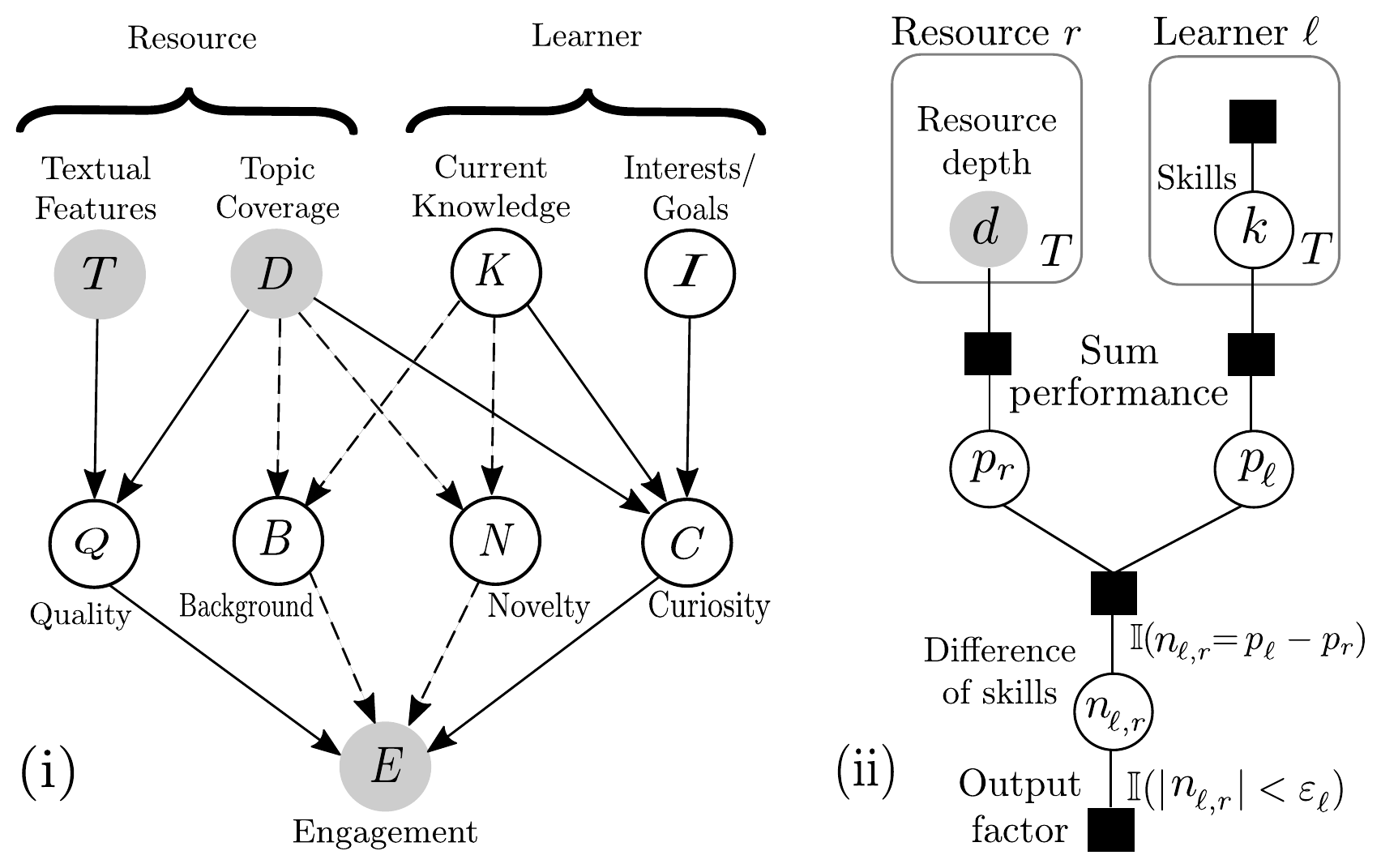}
 \caption{(i) Graphical model representing learner engagement (dashed arrows indicating the components tested) and (ii) TrueLearn factor graph (also, the part with dashed arrows in (i)), integrating resource topics ($d$), current knowledge ($k$) and novelty ($n$) to predict engagement (output factor). $\varepsilon_\ell$ is a dynamic factor of learner $\ell$ indicating the engagement margin with respect to the amount of novelty. Plates represent $T$ top ranked Wikipedia topics.}
 \label{fig:learner_model}
\end{figure}

\section{Our Approach}
We identify four factors that influence learners' engagement and develop a probabilistic graphical model that aims to recover those hidden variables using implicit engagement signals. Using a graphical model that learns from implicit engagement allows us to infer these hidden variables without compromising learner experience through excessive explicit user interventions. The identified factors are: i) baseline resource quality ($Q$), how engaging a resource is for the average learner; ii) background knowledge of the learner ($B$); iii) novelty of the learning material ($N$); and iv) curiosity or learning goals ($C$) of the learner as outlined in \figurename{ \ref{fig:learner_model}}. As a first step, we reformulate the IRT TrueSkill algorithm \cite{trueskill}, to model learner knowledge and novelty as a function of engagement (dashed arrows in \figurename{ \ref{fig:learner_model} (i)}). 

\emph{TrueSkill} has several features that make it an excellent starting point. It is a scalable and online algorithm that shares similarities with our problem and provides a good framework for embedding novelty and a dynamic learner factor (that accounts for knowledge changing over time). TrueSkill algorithm and its successor, TrueSkill 2 \cite{trueskill2}, have been deployed and time-tested with millions of users playing multiplayer video games in the Microsoft Xbox Live system giving substantial evidence of its scalability. The TrueSkill framework also provides a method to address dynamic factor involved in learning how the knowledge state of players changes over time \cite{NIPS2007_3331}. The Gaussian skill parameter in TrueSkill, when used with a humanly interpretable knowledge component space (e.g. the Wikipedia topics covered in a resource), provides an intuitive and transparent knowledge representation. We propose several reformulations of TrueSkill in \cite{Bulathwela2020}, which we name \emph{TrueLearn}. We also propose in \cite{Bulathwela2020} a reformulation of Knowledge Tracing to our problem, demonstrating however in a large dataset the superiority of TrueSkill inspired algorithms. 


\paragraph{Data:} We construct a dataset from the popular video lectures repository VideoLectures.Net (VLN). Since handcrafting the \emph{Knowledge Components (KCs)} in a resource is not scalable, we use an automatic entity linking algorithm, known as Wikification \cite{wikifier}.  The English transcription of the lecture (or the English translation) is used to annotate the lecture with the 5 most relevant knowledge components using a Wikipedia text ontology through Wikifier \cite{wikifier}. This allows us to work with multiple languages and modalities and automatise the extraction of KCs. We divide the lecture text into multiple fragments of approximately 5,000 characters (equivalent roughly to 5 minutes of lecture) before Wikification. The engagement label is computed by calculating the normalised watch time \cite{Guo_vid_prod}. The final dataset consists of 18,933 unique learners.

\paragraph{Models:} We implement four baseline models to compare TrueLearn against: i) \emph{Na\"ive persistence}, which assumes a static behaviour for all users, i.e. if the learner is engaged, they will remain engaged and vice versa; ii) \emph{Na\"ive majority}, which predicts future engagement based solely on mean past engagement of users; iii) KT model (\emph{Multi-Skill KT}) according to \cite{bishopsnewbook}; and iv) \emph{Vanilla TrueSkill} \cite{trueskill}.

\begin{table}[!h]
  \caption{Mean F1-Score with the full VLN dataset}
  \label{tab:truelearn_performance}
  \centering
  \begin{tabular}{l r}  \toprule
    Algorithm 
    & F1-Score \\  \hline
   
    Na\"ive persistence 
    & 0.629 \\
    Na\"ive majority 
    & \textit{0.640} \\
    Vanilla TrueSkill
    & 0.400 \\
    Multi skill KT 
    & 0.259 \\
    TrueLearn 
    & \textbf{0.677} \\  
    \hline
  \end{tabular}
\end{table}

\paragraph{Conclusions:} The results in Table \ref{tab:truelearn_performance} show evidence that \emph{TrueLearn} outperforms the baselines while retaining a transparent learner model. The model is run per learner and trained in an online fashion, thus being scalable. The next step is to model content quality and learner curiosity within the same framework. Exploration into future user interfaces for learning with lecture fragments and ways to planning learning trajectories and recommending material are also timely.

\section{Acknowledgments}
 This research is conducted as part of the X5GON project (www.x5gon.org) funded from the EU's Horizon 2020 research and innovation programme grant No 761758 and partially funded by the EPSRC Fellowship titled "Task Based Information Retrieval", under grant No EP/P024289/1.

\bibliography{aaai}
\bibliographystyle{aaai}
\end{document}